\documentclass{statsoc}     
\usepackage{geometry}      
\usepackage{natbib}		

\usepackage{amssymb}
\usepackage{amsmath}
\usepackage{graphicx}
\usepackage{subfigure}

\usepackage{float}

\title[Curse of Statistics: ``Score-based likelihood ratio'']{Defence Against the Modern Arts: the Curse of Statistics\\  ``Score-based likelihood ratios''}
\author[Neumann and Ausdemore]{C. Neumann and M.A. Ausdemore}
\address{Department of Mathematics and Statistics, South Dakota State University, Brookings, SD, USA.}
\email{Cedric.Neumann@me.com}

\begin{document}

\date{October 2019: Version 1}

\begin{abstract}
For several decades, legal and scientific scholars have argued that conclusions from forensic examinations should be supported by statistical data and reported within a probabilistic framework. Multiple models have been proposed to quantify the probative value of forensic evidence. Unfortunately, several of these models rely on ad-hoc strategies that are not scientifically sound. The opacity of the technical jargon used to present these models and their results, and the complexity of the techniques involved make it very difficult for the untrained user to separate the wheat from the chaff. This series of papers is intended to help forensic scientists and lawyers recognise limitations and issues in tools proposed to interpret the results of forensic examinations. This paper focuses on tools that have been proposed to leverage the use of similarity scores to assess the probative value of forensic findings. We call this family of tools ``score-based likelihood ratios''. In this paper, we present the fundamental concepts on which these tools are built, we describe some specific members of this family of tools, and we explore their convergence to the Bayes factor through an intuitive geometrical approach and through simulations. Finally, we discuss their validation and their potential usefulness as a decision-making tool in forensic science. 
\end{abstract}


\pagebreak

\section{Introduction}\label{intro}


For more than half a century, legal and scientific scholars have widely advocated Bayesian reasoning for handling the uncertainty in the determination of the source of forensic evidence (see \cite{Evett1998} for a historical reference, \cite{AitkenTaroni:2004} for a general introduction). Bayesian inference revolves around the use of the Bayes factor to update one's prior beliefs about two competing propositions related to the source of the evidence. The updated beliefs are often called posterior beliefs. Posterior beliefs are probabilities and do not equate to categorical decisions. The path leading from a posterior probability to a decision involves the use of loss functions and has been described, in the forensic context, by \cite{Biedermann2008}. Proponents of Bayesian reasoning argue that it is the only coherent and logical manner for performing inferences in forensic science. They further argue that, in casework, forensic scientists do not possess the information that would allow them to assign prior beliefs to the propositions that are considered. Consequently, forensic scientists should limit themselves to reporting Bayes factors and let others (e.g., fact-finders, jurors, judges) complete the inference process. Therefore, the challenge for forensic scientists is to assign values to the Bayes factors for various evidence types (e.g., fibre, paint, glass, fingerprints, footwear impressions, handwriting, toolmarks, etc.).

Forensic scientists have been able to assign Bayes factors to simple forms of forensic evidence for many years\footnote{The adjective ``simple'' refers to the level of complexity of the mathematical representation of the evidence and of the probabilistic models involved; it is not used to qualify how the evidence is transferred, recovered or analysed.}. For example, the statistical models used to quantify the weight of single DNA profiles or simple mixtures of DNA profiles are well understood\footnote{A single DNA profile is usually represented by a small set of independent bivariate categorical vectors, which joint distributions under the two competing propositions are usually trivial to model.}. Conversely, only anecdotical attempts have been made to assign Bayes factors to complex forms of forensic evidence, such as handwriting and fingerprint evidence \citep{Bozza2008, Forbes2014, Neumann2015, Tackett2018}. 

Assigning Bayes factors to complex evidence forms requires defining reasonable likelihood functions to represent the joint distributions of heterogenous and high-dimensional feature vectors\footnote{For example, in the case of fingerprint evidence, a single minutia can be represented by its Euclidean coordinates (bivariate continuous variable), its type (nominal variable) and its direction in the ridge flow (circular variable). An impression where $n$ minutiae are observed is then represented by a $4n$-long vector, which contains three different types of variables.}.

To bypass the need to work with intractable likelihood functions, researchers have concentrated on the use of \textit{(dis)similarity metrics} or \textit{kernel functions} to reduce the complexity and dimensionality of the problem. These attempts have given rise to a family of ad-hoc methods aimed at describing the probative value of forensic evidence. We call these methods ``score-based likelihood ratios''. 

In this paper, we show that these ad-hoc tools offered to support Bayesian inference of the source of complex forms of evidence may have some merits as deterministic decision tools; however, their use within a Bayesian paradigm is not appropriate. As a result, they cannot be used to update prior beliefs on the source of a finger impression as part of Bayesian reasoning, and they are not fulfilling the requirements set forth by the legal and scientific scholars who advocate for a move towards a more formal Bayesian approach in forensic science. Specifically, we show that:
\begin{enumerate}
	\item Some tools do not address the question of interest;
	\item Some tools can induce incoherent inference, in the sense that they can support both mutually exclusive propositions using the same information obtained from the evidence; 
	\item Some tools may unpredictably over- or underestimate the weight of the evidence represented by a set of trace and control objects. 
\end{enumerate}

\section{Common source vs. specific source scenarios}\label{Hypotheses}

The inference of the identity of the donor of a trace from its comparison with control material from a known source requires considering two mutually exclusive hypotheses, denoted $H_0$ and $H_1$ below\footnote{These hypotheses are commonly called the \textit{prosecution hypothesis} and \textit{defence hypothesis} \citep{AitkenTaroni:2004}.}. A certain lack of formalism in the formulation of these hypotheses has resulted in the development of models and the collection of data that mismatch the needs of the criminal justice system.

The next sections briefly develop two formal scenarios that frame the inference of the source of forensic evidence: the \textit{common source scenario} and the \textit{specific source scenario} \citep{Ommen:2017}. These scenarios are often confused with one another. This results in the development of models under one scenario to answer the question considered by the other one. Thus, understanding their differences is important and helps assess the potential and limitations of the different inference frameworks for forensic evidence. 

\subsection{Common source scenario}\label{CS.Hypotheses}

The common source scenario considers whether two pieces of forensic evidence originate \textit{from the same source} or from \textit{different sources} without formally specifying which sources are considered. This scenario typically relates to inference of the source of two trace samples, $e_{u_1}$ and $e_{u_2}$  (e.g., two finger impressions recovered on two different crime scenes or even on the same crime scene), with the goal of determining if they originate from the same unknown source (e.g., determining whether the two scenes are linked or whether there were one or more perpetrators). 

The hypotheses considered in the common source scenario can be stated as follows:

\begin{itemize}
	\item[] $H_{0_{CS}}$: $e_{u_1}$ and $e_{u_2}$ originate from the same, unknown, source; 
	\item[] $H_{1_{CS}}$: $e_{u_1}$ and $e_{u_2}$ originate from two different, unknown, sources.
\end{itemize}

In this scenario, the true source of each piece of evidence is considered to be a random source from a population of potential sources. Under $H_{0_{CS}}$, the source of the two pieces of evidence is the same random source, while the evidence material originate from two different random sources under $H_{1_{CS}}$

\subsection{Specific source scenario}\label{SS.Hypotheses}

Contrary to the previous scenario, the specific source scenario typically involves the comparison of trace material, $e_{u}$, with control material from a known source, $e_s$, with the goal of determining if the trace was made by the considered source. The hypotheses considered in the specific source scenario can be stated as follows:

\begin{itemize}
	\item[] $H_{0_{SS}}$: $e_{u}$ and $e_{s}$ were made by Source X.; 
	\item[] $H_{1_{SS}}$: $e_{u}$ was made by another source than Source X.
\end{itemize}

In this scenario, Source X is identified. It can be considered fixed. Under $H_{1_{SS}}$, the true source of $e_{u}$ is unknown and is considered to be a random source from a population of potential sources, while Source X remains the undisputed donor of $e_{s}$.

The distinction between both scenarios is not merely theoretical. Each scenario results in different likelihood functions for the same information, and in different interpretations of the results of forensic examinations. 

In the vast majority of cases, the inference questions of greatest interest to the criminal justice system fall under the umbrella of the specific source scenario. Nevertheless,  the determination that two pieces of evidence were made by the same unknown source may be relevant to some investigations (e.g., for forensic intelligence-led investigations). Since these two scenarios are different and consider two radically different pairs of hypotheses, it seems intuitive that they should not be interchanged. Unfortunately, they are often confused.

\subsection{Generative models}\label{Generative.models}

This paper explores the convergence of different models and inference frameworks partly through simulations. The simulations rely on generative models that give simplified representations of how the data arise under the different hypotheses laid out in Sections \ref{CS.Hypotheses} and \ref{SS.Hypotheses}. These simplified models are used throughout the paper and are introduced below. 

We consider a simple univariate setting to explore the construction and convergence of the different Bayes factors in the common and specific source scenarios. The common source scenario considers whether two traces originate from the same, unknown, source; thus, the generative models under both common source hypotheses can be represented by two hierarchical random effects models:

\begin{itemize}
\item[]	$e_{u_1} = \mu + d_1 + u_1$, where $d_1 \sim N(0,\sigma^2_d)$ and $u_1 \sim N(0,\sigma^2_{u_1})$;
\item[]	$e_{u_2} = \mu + d_2 + u_2$, where $d_2 \sim N(0,\sigma^2_d)$ and $u_2 \sim N(0,\sigma^2_{u_2})$;
\end{itemize}
where $\mu$ is the mean of the population of sources, $d_1$ and $d_2$ are random effects due to sources, and $u_1$ and $u_2$ are random effects due to objects within sources\footnote{If we consider the practical example of fingerprint evidence, $\mu$ represents the mean of the distribution of the characteristics of all friction ridge skin in a population; $d_1$ and $d_2$ represent the deviations between the overall mean of the population, $\mu$, and the friction ridge characteristics of the first and second sources; $u_1$ and $u_2$ are random effects that affect the final appearance (after development, transfer, photography, etc.) of fingerprints resulting from different impressions of the fingers represented by $u_1$ and $u_2$ on various surfaces. The effects $u_1$ and $u_2$ may be distinct as two impressions may be affected by different sets of factors}. 

Under $H_{0_{CS}}$, the two pieces of evidence originate from the same source and, thus, have the same value for $d_1$ and $d_2$\footnote{Note that they do not necessarily have the same value for $\sigma^2_u$ if the different pieces of evidence were left under different conditions.}. Under $H_{1_{CS}}$, the two pieces of evidence originate from two different sources and are therefore independent. Thus, the respective joint distributions of $e_{u_1}$ and $e_{u_2}$ are:

\begin{eqnarray}
	\begin{pmatrix} e_{u_1} \\ e_{u_2} \end{pmatrix}	| H_{0_{CS}} & \sim & MVN \left(\begin{pmatrix} \mu \\ \mu \end{pmatrix}, \begin{pmatrix} \sigma^2_d + \sigma^2_{u_1} & \sigma^2_d \\  \sigma^2_d & \sigma^2_d + \sigma^2_{u_2} \end{pmatrix} \right);\nonumber \\	
	\begin{pmatrix} e_{u_1} \\ e_{u_2} \end{pmatrix}	| H_{1_{CS}} & \sim & MVN \left(\begin{pmatrix} \mu \\ \mu \end{pmatrix}, \begin{pmatrix} \sigma^2_d + \sigma^2_{u_1} & 0 \\  0 & \sigma^2_d + \sigma^2_{u_2} \end{pmatrix} \right).
	\label{CS.generative.model}
\end{eqnarray}

The generative models in the specific source scenario differ depending on whether $H_{0_{SS}}$ or $H_{1_{SS}}$ is considered. Under $H_{0_{SS}}$, when all evidence originate from the same source, the models are two simple random effects models:
 
\begin{itemize}
\item[] $e_u = \mu_d + u$, where $u \sim N(0,\sigma^2_u)$;
\item[] $e_s = \mu_d + s$, where $s \sim N(0,\sigma^2_s)$;
\end{itemize}
and, where $\mu_d$ represents the mean for the considered specific source, and $u$ and $s$ are random effects respectively corresponding to trace and control samples. 

Under $H_{1_{SS}}$, the generative model for the control material from the specific source is the same as under $H_{0_{SS}}$ (indeed, there is no dispute that $e_s$ originates from the known source). However, the model for the trace material, $e_u$, is a hierarchical random effects model to reflect that its true source is an unknown source in the population of potential sources:  

\begin{itemize}
\item[] $e_u = \mu + d + u$, where $d \sim N(0,\sigma^2_d)$ and $u \sim N(0,\sigma^2_u)$;
\item[] $e_s = \mu_d + s$, where $s \sim N(0,\sigma^2_s)$;
\end{itemize}
and where $\mu$, $\mu_d$, $d$, $u$ and $s$ are defined as above\footnote{A similar analogy to the one made in Footnote 5 can be made here. The constant $\mu_d$ represents the characteristics of the friction ridge skin of a specific finger from a known individual (e.g., a suspect). The effect $d$ represents the characteristics of the friction ridge skin of a specific finger from an unknown individual (e.g., the true donor of the latent print). The random effects $u$ and $s$ affect the final appearance (after development, transfer, photography, etc.) of fingerprints resulting from different impressions of the fingers represented by $\mu_d$ and $d$ on various surfaces. The variance terms, $\sigma^2_u$ and $\sigma^2_s$, may be distinct as latent and control prints are acquired under different sets of conditions.}.

Under $H_{0_{SS}}$, trace and control materials are independent given $\mu_d$, and their joint distribution is multivariate normal. Under $H_{1_{SS}}$, trace and control materials are independent since they are not from the same source, and their joint distribution is also multivariate normal. We have:

\begin{eqnarray}
	\begin{pmatrix} e_u \\ e_s \end{pmatrix}	| H_{0_{SS}} & \sim & MVN \left(\begin{pmatrix} \mu_d \\ \mu_d \end{pmatrix}, \begin{pmatrix} \sigma^2_u & 0 \\  0 & \sigma^2_s \end{pmatrix} \right);\nonumber \\
	\begin{pmatrix} e_u \\ e_s \end{pmatrix}	| H_{1_{SS}} & \sim & MVN \left(\begin{pmatrix} \mu \\ \mu_d \end{pmatrix}, \begin{pmatrix} \sigma^2_d + \sigma^2_u & 0 \\  0 & \sigma^2_s \end{pmatrix} \right).
	\label{SS.generative.model}
\end{eqnarray}

If we take the view that forensic evidence must be evaluated within a Bayesian paradigm, then we are interested in quantifying the weight of the evidence using Bayes factors (or, when the parameters are known, likelihood ratios). In the common source framework, the likelihood ratio for $e_{u_1}$ and $e_{u_2}$ is \citep{Ommen:2017}: 
\begin{equation}
	LR_{CS} = \frac{f(e_u,e_s|H_{0_{CS}})}{f(e_u,e_s|H_{1_{CS}})}=\frac{f(e_u,e_s|H_{0_{CS}})}{f(e_u|H_{1_{CS}})f(e_s|H_{1_{CS}})};
	\label{CS.LR}
\end{equation}
while the likelihood ratio for $e_{u}$ and $e_{s}$ in the specific source framework is \citep{Ommen:2017}: 
\begin{equation}
	LR_{SS} = \frac{f(e_u,e_s|H_{0_{SS}})}{f(e_u,e_s|H_{1_{SS}})} = \frac{f(e_u|H_{0_{SS}})}{f(e_u|H_{1_{SS}})}.
	\label{SS.LR}
\end{equation}

\subsection{Convergence of specific and common source Bayes factors}
\label{Convergence.SS.CS}

We already mentioned that, in many cases, forensic scientists are working within the context of the specific source scenario. They are provided with trace material and they want to infer whether it originates from the same specific source that was used to obtain the control material. Using the toy examples in Equations (\ref{CS.generative.model}) and (\ref{SS.generative.model}), we can study the convergence of the common source likelihood ratio in Equation (\ref{CS.LR}) to the specific source likelihood ratio in Equation (\ref{SS.LR}) that should be used to quantify appropriately the weight of the evidence.

To compare these likelihood ratios, we consider pairs of $e_u$ and $e_s$ generated by the model in Equation (\ref{SS.generative.model}) under $H_{0_{SS}}$ or $H_{1_{SS}}$ and we calculate the likelihood ratios in Equations (\ref{CS.LR}) and (\ref{SS.LR}). To calculate the common source likelihood ratio using the data generated under the specific source model, we set $e_{u_1} = e_u$, $e_{u_2} = e_s$, $\sigma^2_{u_1} = \sigma^2_u$ and $\sigma^2_{u_2} = \sigma^2_s$. 

\begin{figure}[H]
\begin{center}
\subfigure[]{\includegraphics[bb=0in 0in 11in 5.2in, scale=0.35]{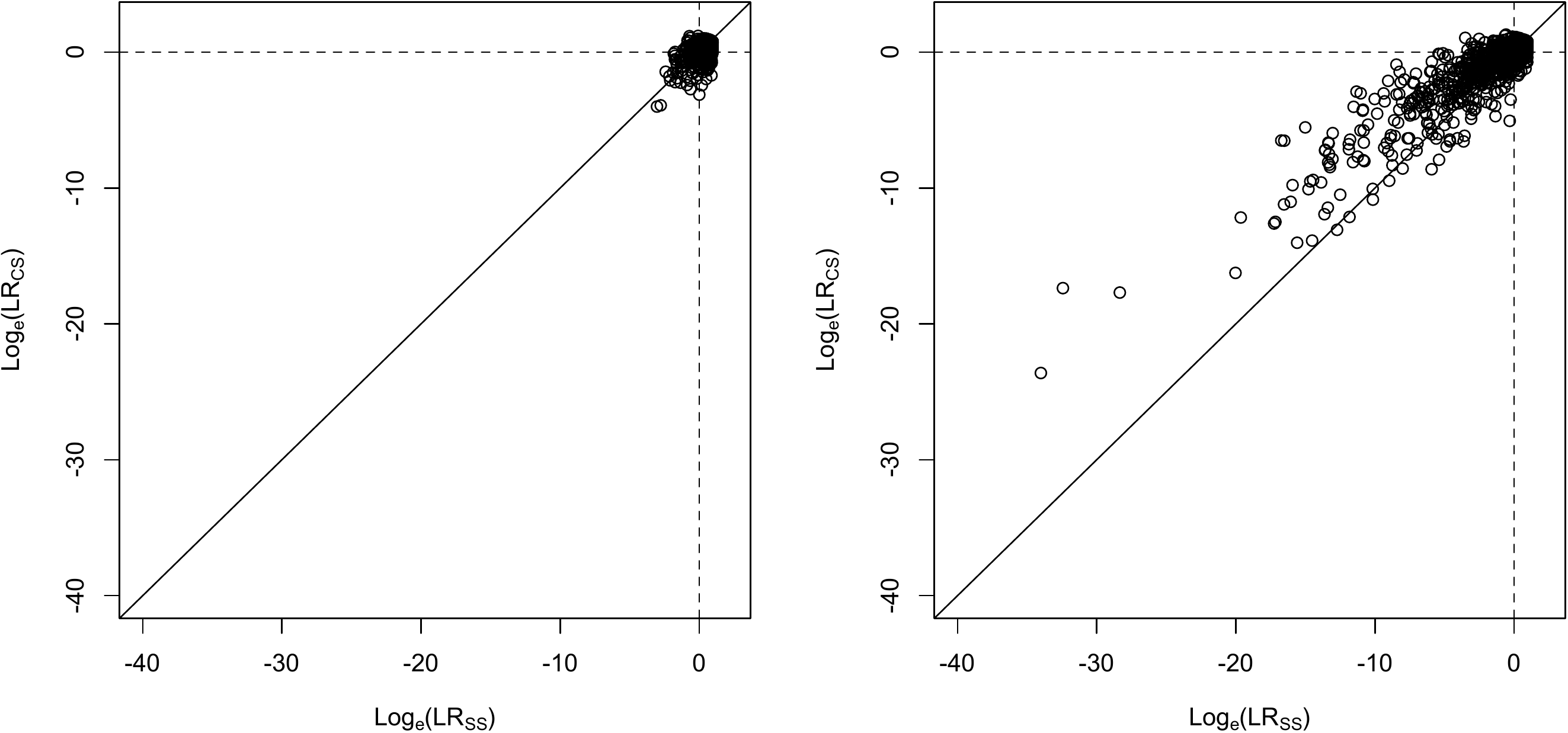}}
\subfigure[]{\includegraphics[bb=0in 0in 11in 5.2in, scale=0.35]{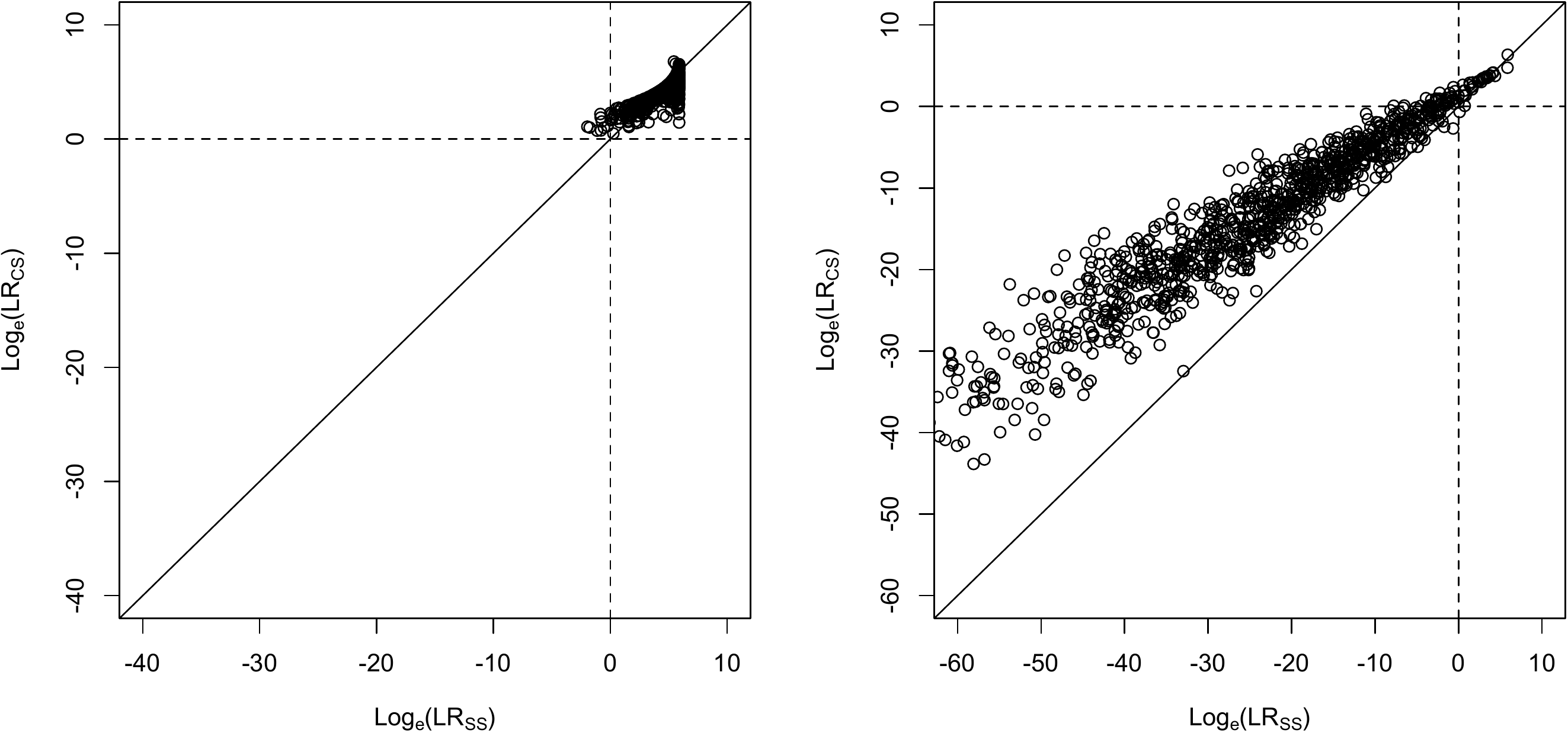}}
\subfigure[]{\includegraphics[bb=0in 0in 11in 5.2in, scale=0.35]{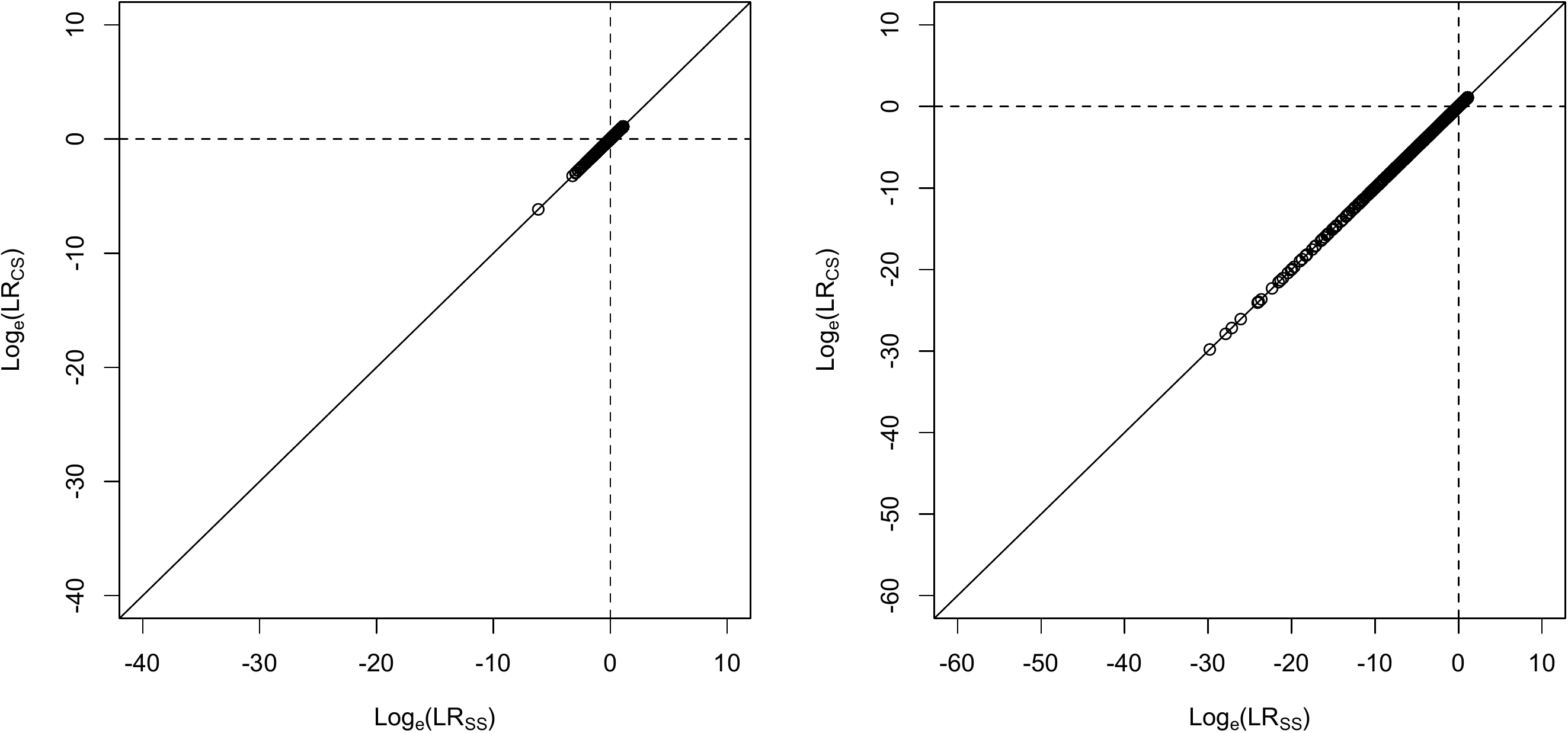}}
\end{center}
\caption{Comparisons between LRs in the common source (y-axis) and specific source scenarios (x-axis). Columns: the left column reports the results when $e_u$ and $e_s$ have been sampled under $H_{0_{SS}}$; the right column reports the results under $H_{1_{SS}}$. Rows: (a) the source of the control material has common characteristics; (b) the source of the control material has rare characteristics; (c) the source of the control material has common characteristics, however it has virtually no variance.}
\label{CS.vs.SS.LR}
\end{figure}

Figure \ref{CS.vs.SS.LR} presents the results of three experiments. In all three experiments, $\mu=10$, $\sigma^2_d=10$ and $\sigma^2_u=2$. All simulations were repeated 1,000 times. In the first experiment, the characteristics of the source of $e_s$ were chosen to be relatively common with respect to the population of sources ($\mu_d=9$) but also quite variable ($\sigma^2_s=1$). In the second experiment, the characteristics of the source of $e_s$ were chosen to be rare with respect to the population of sources ($\mu_d=0$) but remained variable ($\sigma^2_s=1$). In the last experiment, the variability of the characteristics of the known source of the control material was chosen to be virtually negligible ($\sigma^2_s=10^{-5}$)\footnote{The values for the models' parameters are chosen to represent different situations: paint, glass and fibres are material with somewhat large within-source variability and their specificity is variable; control finger impressions have low within-source variability and contain a large number of very discriminative features; finally, DNA profiles obtained directly from individuals have virtually not within-source variability (in terms of allelic designation) and are highly specific to these individuals.}.

The results of the experiments show that likelihood ratios for the common and the specific source scenarios do not converge unless the variability of the source of the control material is negligible\footnote{This is typically the case for forensic DNA analysis when single full DNA profiles are considered. Since the allelic designation of a full DNA profile is extremely reproducible, the inference of the identity of source of a pair of full DNA profiles will be the same under both common and specific source scenarios. This may explain why the distinction between common and specific source scenarios was not discussed until recently by \cite{Ommen:2017}.}. Importantly, the results for the first two experiments in Figure \ref{CS.vs.SS.LR} show that the common source likelihood ratio unpredictably over- or underestimates the value of the specific source likelihood ratio. That said, while assigning a common source likelihood ratio when $H_{1_{SS}}$ is true may underestimate the corresponding specific source likelihood ratio, Figure \ref{CS.vs.SS.LR}  shows that common source likelihood ratios have a marked tendency to overestimate their counterparts (we have not found a situation where common source likelihood ratios consistently underestimate specific source likelihood ratios). 

The lack of convergence raises issues regardless of whether $H_{0_{SS}}$ or $H_{1_{SS}}$ is true. When $H_{0_{SS}}$ is true, underestimating the value of the specific source likelihood ratio may result in the erroneous exclusion of the source of the control impressions as the source of the trace impression. While this is an obvious issue, the criminal justice system currently considers this to be a better outcome than the erroneous identification of an innocent. Furthermore, when $H_{0_{SS}}$ is true, overestimating the value of the specific source likelihood ratio only results in being overconfident in the support of the correct conclusion that the source of the control material is also the source of the trace material; thus, the impact of the overestimation may be considered minimal. Unfortunately, when $H_{1_{SS}}$ is true, overestimating the value of the specific source likelihood ratio may result in exculpatory evidence not being given the appropriate weight in favour of an innocent, yet suspected, source. In fact, Figures \ref{CS.vs.SS.LR}(a) and (b) show that some pieces of evidence result in values of the specific source likelihood ratios that are less than one and values of the common source likelihood ratios that are greater than one. 

Ultimately,  miss-specifying the interpretation framework results in answering the wrong question, and may result in serious miscarriages of justice when the common source likelihood ratio is used instead of the specific source one.   

\section{Score-based likelihood ratios}
\label{SLR.sec}

The use of scores to calculate score-based likelihood ratios can be tracked back to the late 1990s and early 2000s and the field of speaker recognition, fingerprint and other types of evidence \citep{Meuwly2001, Champod2001.Ear, Gonzalez-Rodriguez:2003, Gonzalez-Rodriguez:2005, Gonzalez:2006, Egli:2006, Meuwly2006,  Neumann:2007, Neumann2009.inks, Egli2014}. The natural proximity of these forensic sub-disciplines and biometry have led researchers to quickly realise the benefits of modelling the (dis)similarity between pairs of observations, rather than modelling complex feature vectors in their original space. This enabled them to bypass the need to work with the intractable likelihood functions associated with complex forms of pattern and trace evidence, and, instead, allowed them to model univariate continuous data. 

\begin{figure}[H]
\begin{center}
\includegraphics[width=\textwidth]{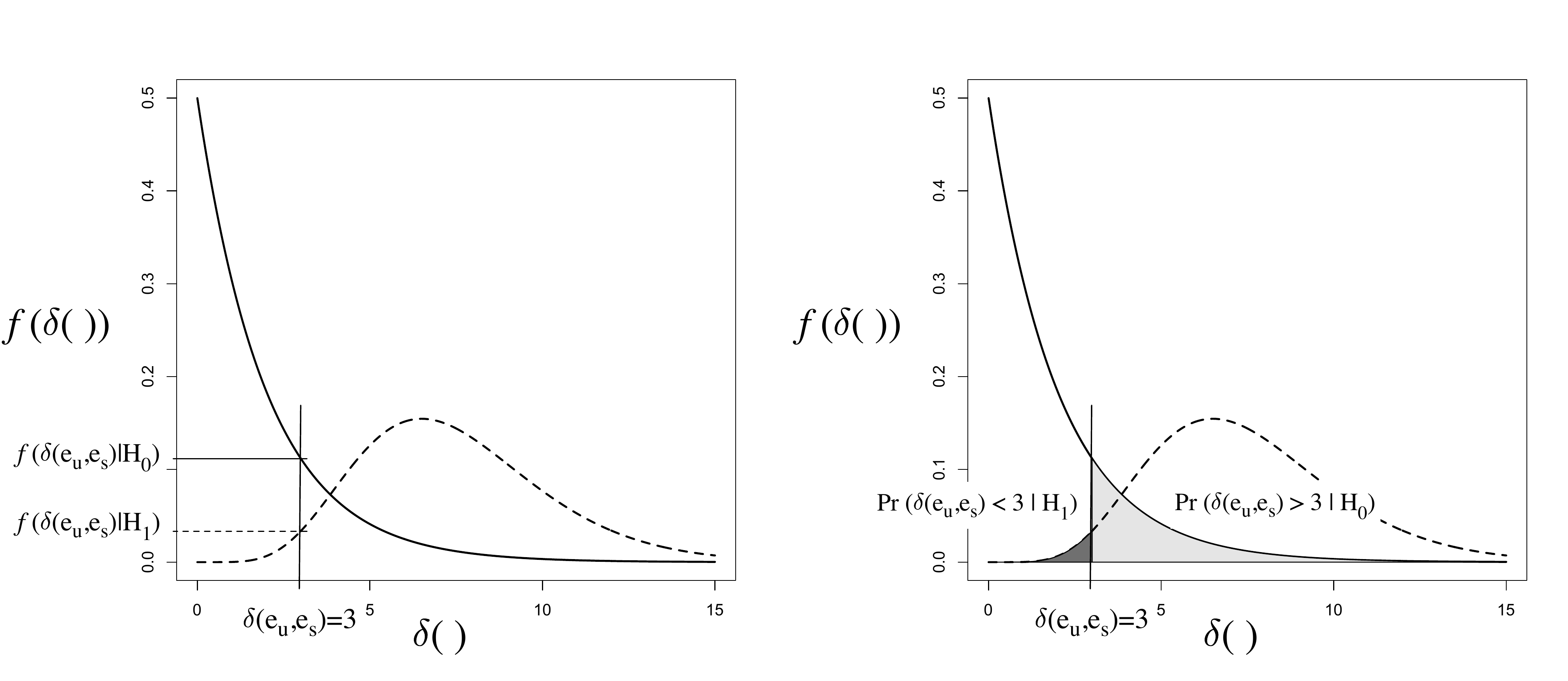}
\caption{Comparison between two different concepts for the use of summary statistics/kernel functions to provide some information on the probative value of fingerprint evidence. Left panel: score-based likelihood ratio obtained by calculating the ratio of the density of a summary statistic for an observed pair of trace and control objects, $\delta(e_u,e_s)$, in its sampling distribution under the first proposition and in its sampling distribution under the second proposition. Right panel: FRStat-like \citep{FRSTAT2018,Neumann2019} ratio obtained by calculating the ratio of $\alpha$- and $\beta$-error types associated with a decision of identification or exclusion at an observed level of dissimilarity of a pair of trace and control impressions, $\delta(e_u,e_s)$, using the two sampling distributions under the two propositions.}
\label{Score.vs.ABC.vs.FRStat}
\end{center}
\end{figure}

Different constructions of score-based likelihood ratios have been proposed over the years and, despite their limitations, their use in casework is advocated (at least in Europe by the \cite{ENFSI2016}). The concept behind most models rests on the comparison of the likelihood of the score calculated between a single trace object, $e_u$, and a single control object from a known source, $e_s$, evaluated in two different density functions. These density functions are based on the sampling distributions of the score under two mutually exclusive propositions. The concept is illustrated in the left panel of Figure \ref{Score.vs.ABC.vs.FRStat} which shows the ratio of $f(\delta(e_u,e_s)|H_0)$ and $f(\delta(e_u,e_s)|H_1)$, where $\delta(e_u,e_s)$ is the score between the single trace and single control objects, and $f(\cdot|H_0)$ and $f(\cdot|H_1)$ represents the sampling distributions of interest. We warn the reader that other models involving scores have been proposed \citep{Armstrong:2017, FRSTAT2018, Ausdemore.2stages.2019, Hendricks2019ABC} but are not considered to be score-based likelihood ratios. These models do not rely on the ratio of the likelihoods of the score in two sampling distributions. For example, the right panel of Figure \ref{Score.vs.ABC.vs.FRStat} shows the concept underlying a model called FRStat \citep{FRSTAT2018}\footnote{See \cite{Neumann2019} for a critic of FRStat}, which relies on the ratio of two tail probabilities bounded by the score between a trace and a control object. 

\subsection{Similarity metrics and kernel functions}\label{scores.and.kernels}

A score can have two interpretations: it can be seen as a summary statistic resulting from the comparison of two objects, or it can be seen as the scalar projection resulting from the inner product of two vectors. In the first case, we talk about \textit{(dis)similarity metrics}, while in the second case we talk about \textit{kernel functions}. Both functions map complex random vectors from their natural space to the real line, $\mathbb{R}$, and both offer great flexibility to researchers. First, researchers can design algorithms that measure the distance between two objects, such that the value representing that distance is minimised when the two objects originate from the same source, and is maximised when they originate from different sources\footnote{Some algorithms maximise the value of the score when the objects originate from the same source, and minimise it when the objects originate from different sources. The upcoming discussion on the use of scores is not affected by their interpretation as \textit{similarity scores} or \textit{distances}.}. Secondly, the level of (dis)similarity between pairs of objects can be expressed as a univariate continuous random variable, which probability distribution is significantly more convenient to model than the distribution of original vectors representing the observations made on the impressions. However, the benefits of being able to work in a continuous univariate space are not without limitations, which are explored below. 

When the function used to calculate scores is considered as a summary statistic, we can discuss the sampling distributions of the score under various situations. When the score function is considered to be a kernel function, the score has a geometric interpretation. Formally, a score interpreted as a summary statistic of the (dis)similarity between two objects $e_i$ and $e_j$ can be defined as $\delta(e_i, e_j)$, where $\delta$ is any function with a real-valued output. A score interpreted as the inner product of two vectors can be similarly defined as $\kappa(e_i, e_j) = \langle \eta(e_i),\eta(e_j) \rangle$, where $\kappa$ is a kernel function, $\eta$ is a set of basis expansions and $\langle \cdot, \cdot \rangle$ is the inner product. The main difference between $\delta$ and $\kappa$ is that $\kappa$ has to be a positive semi-definite symmetric function, while there is no requirement for the construction of $\delta$. 

These two different perspectives on the score function are used to investigate the different score-based models in the next sections. 
The generative models described in Section \ref{Generative.models} are used to discuss the convergence of these models to the specific source likelihood ratio of interest in Equation (\ref{SS.LR}) as we did in Section \ref{Convergence.SS.CS}. In order to perform these simulations, both $\delta$ and $\kappa$ are defined as the squared Euclidean distance, which is both a summary statistic and a valid kernel function, and which also has tractable distributions for the chosen generative models \citep{Helper2012}. 

\subsection{Common source score-based models}\label{CS.SLR.section}

The models in the following sections are best introduced through experiments that allow to study the sampling distributions under the two alternative propositions.

The first type of score-based model is based on results obtained in biometry (Section 4.3 in \cite{Ross:2006})\footnote{Early papers on the use of scores to approximate Bayes factors in forensic science lack clarity on how the sampling distributions of the scores were studied. While it seems that the work by \cite{Champod2001.Ear} and \cite{Gonzalez-Rodriguez:2005} describes asymmetric score-based likelihood ratios (see Section \ref{AsySLR}), it may be that they are in fact common source score-based likelihood ratios (or at least calculated as such).}. The score, $\delta(e_u,e_s)$, is evaluated using sampling distributions based on the following thought experiments:
\begin{enumerate}
	\item When the prosecutor proposition is correct, $\delta(e_u,e_s)$ is a score that is calculated by comparing trace and control material from the same, random, source. The sampling distribution of $\delta(e_u,e_s)$ under $H_0$ can be studied by considering a sample of sources from a relevant population, and by sampling and comparing a single trace and a single control object from each source.
	\item When the defence proposition is correct, $\delta(e_u,e_s)$ is a score that is calculated by comparing trace and control objects from different sources. The sampling distribution of $\delta(e_u,e_s)$ under $H_1$ can be studied by sampling independent pairs of objects from a relevant population, and by comparing a trace object from the first source to a control object from the second source.
\end{enumerate}

This type of model has one main advantage: both sampling distributions can be learned ahead of time based on a large sample of sources from a relevant population. Once learned, the two sampling distributions can be used for any new case. It also has two main limitations. First, it is only reporting the \textit{average} density of $\delta(e_u,e_s)$ under both propositions. Neither sampling distribution is specific to the donor of $e_s$. This type of model clearly addresses the common source pair of propositions and is not relevant to a specific case involving the comparison of a trace object to known control material from a given source. Second, Bayes factors can roughly be viewed as the ratio between some measure of \textit{similarity} between the characteristics of the trace and control objects, and some measure of the \textit{rarity} of the characteristics of the trace. However, the model described above does not account for the rarity of the trace characteristics at all. This type of model only accounts for the rarity of the level of similarity\footnote{For example, consider a bloodstain recovered at a crime scene. A suspect, who has blood of the same type as the bloodstain, is considered. Clearly, the information that the type of the blood recovered at a crime scene is the same as the one of the suspect will be a lot more helpful to support the inference that the blood comes from the suspect if the blood type is AB$^-$ (less than 1\% of the population) than if the blood type is O$^+$ (approx. 40\% of the population). Yet, under the defence proposition, the model described above only assigns a probability to the event that the two blood types correspond by chance without accounting for the specific type of the blood at the crime scene.}. 

To compare this type of model to the specific source likelihood ratio in Eq. \ref{SS.LR}, we use the generative models proposed in Eq. \ref{CS.generative.model}. By defining $\delta(e_u,e_s) = (e_{u_1} - e_{u_2})^2$, we have that: 
\begin{eqnarray}
	(e_{u_1} - e_{u_2})^2 | H_{0_{CS}}  & \sim & \frac{1}{\sigma^2_{u_1} + \sigma^2_{u_2}}\chi^2 \left( \frac{(e_{u_1} - e_{u_2})^2}{\sigma^2_{u_1} + \sigma^2_{u_2}} \right),\nonumber \\
	(e_{u_1} - e_{u_2})^2 | H_{1_{CS}}  & \sim & \frac{1}{\sigma^2_{u_1} + \sigma^2_{u_2} + 2\sigma^2_p}\chi^2 \left( \frac{(e_{u_1} - e_{u_2})^2}{\sigma^2_{u_1} + \sigma^2_{u_2} + 2\sigma^2_p} \right),
\end{eqnarray}
and that:
\begin{equation}
	SLR_{CS} = \frac{f((e_{u_1} - e_{u_2})^2|H_{0_{CS}})}{f((e_{u_1} - e_{u_2})^2|H_{1_{CS}})}.
	\label{CS.SLR}
\end{equation}

\begin{figure}[H]
\begin{center}
\subfigure[]{\includegraphics[bb=0in 0in 11in 5.2in, scale=0.35]{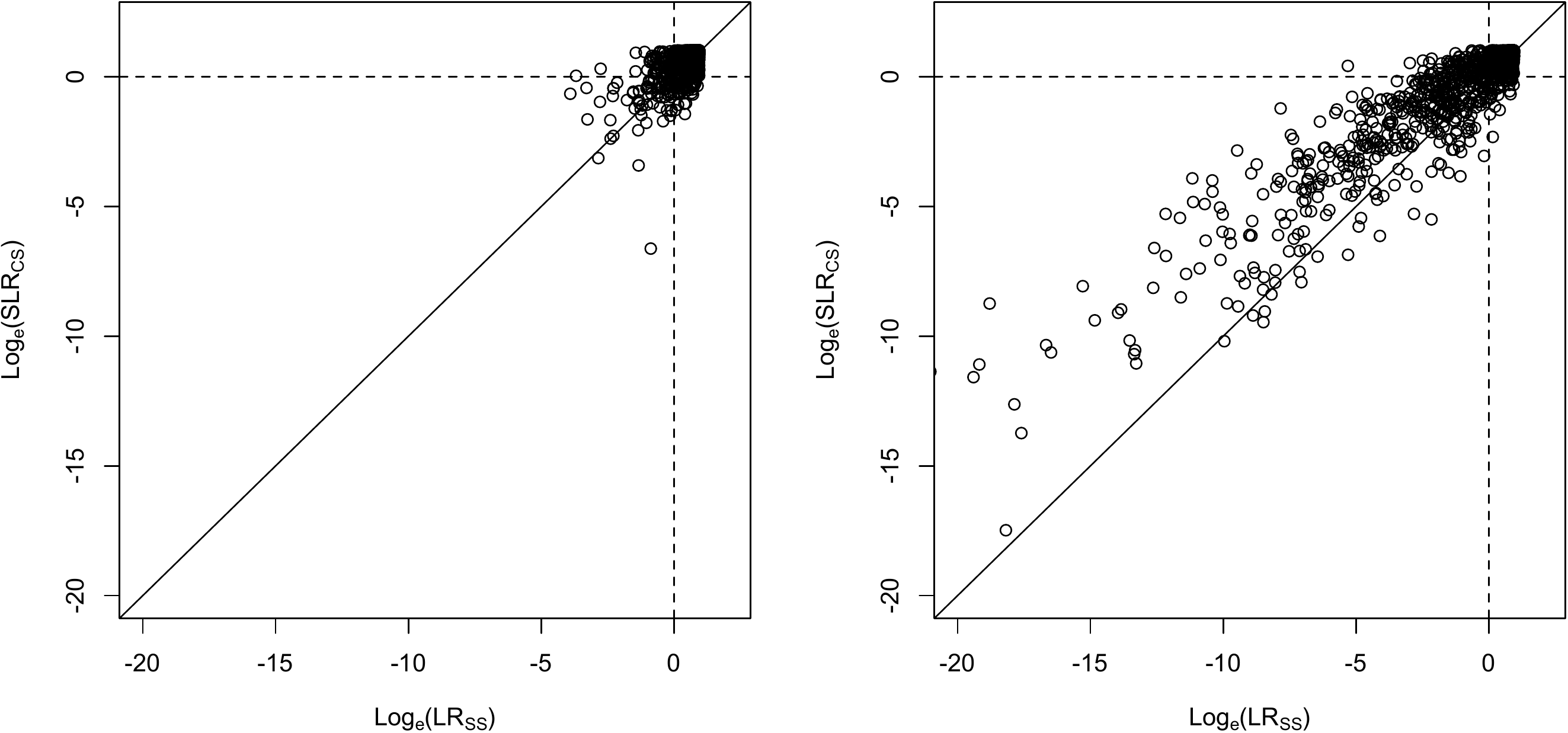}}
\subfigure[]{\includegraphics[bb=0in 0in 11in 5.2in, scale=0.35]{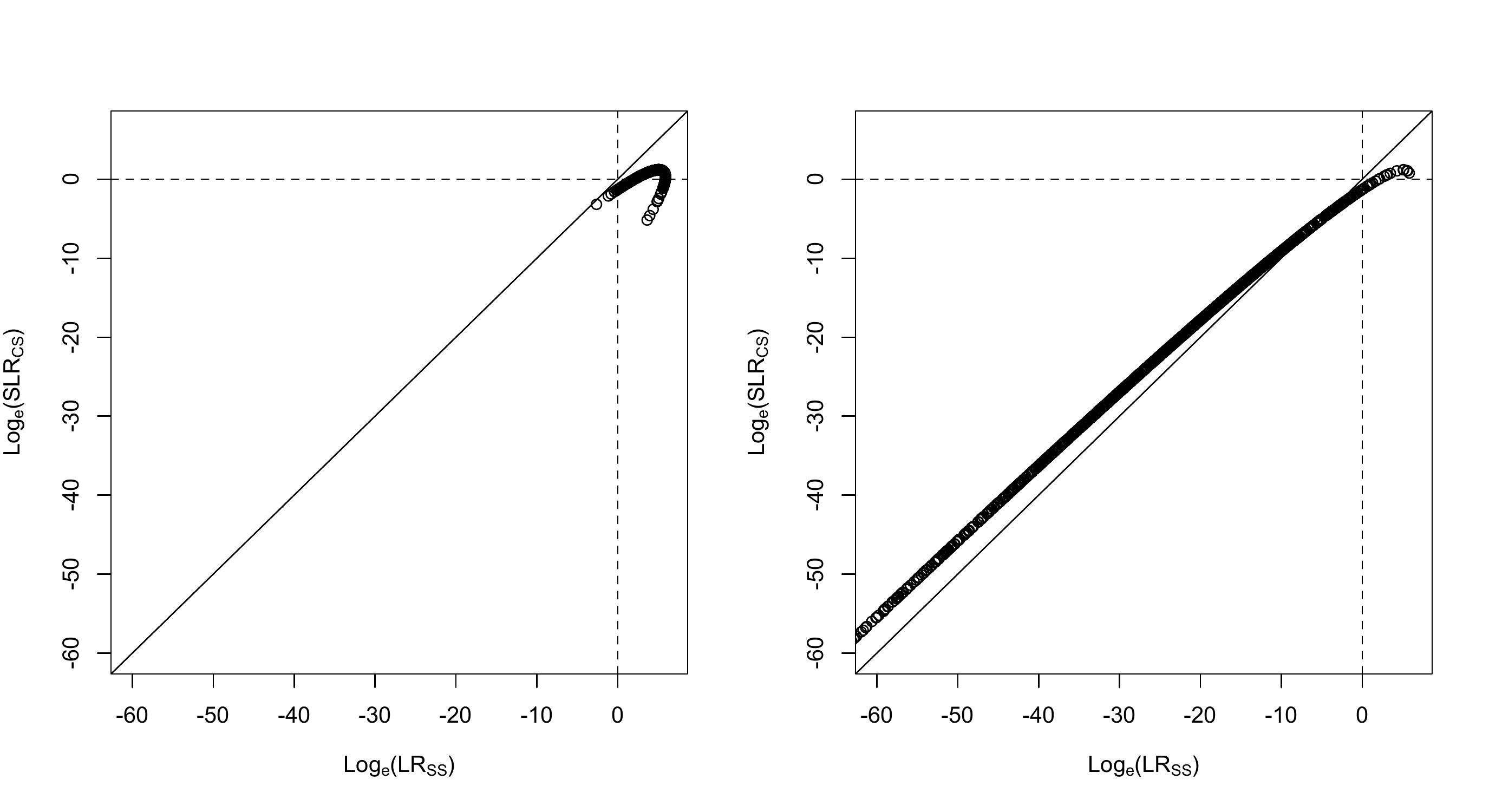}}
\end{center}
\caption{Comparisons between SLRs in the common source scenario with the LR in the specific source scenario. Columns: the left column reports the results when $e_u$ and $e_s$ have been sampled under $H_{0_{SS}}$; the right column reports the results under $H_{1_{SS}}$. Rows: (a) the source of the control impression is common and has some variance; (b) the source of the control impression is rare and has virtually no variance.}
\label{CS.SLR.vs.SS.LR}
\end{figure}

The results of the comparison of Equations (\ref{SS.LR}) and (\ref{CS.SLR}) using our toy example are presented in Figure \ref{CS.SLR.vs.SS.LR}. To study the convergence of $LR_{SS}$ and $SLR_{CS}$, we set $e_{u_1} = e_u$, $e_{u_2} = e_s$, $\sigma^2_{u_1} = \sigma^2_u$ and $\sigma^2_{u_2} = \sigma^2_s$. In both models, $\mu=10$, $\sigma^2_d=10$ and $\sigma^2_u=2$. All simulations were repeated 1,000 times. In the first experiment, the characteristics of the donor of $e_s$ were chosen to be relatively common with respect to the population of donors ($\mu_d=9$) but also quite variable ($\sigma^2_s=1$). In the second experiment, the characteristics of the donor of $e_s$ were chosen to be rare with respect to the population of donors ($\mu_d=0$) with a virtually negligible variability ($\sigma^2_s=10^{-5}$). 

The same general conclusions drawn from the data presented in Figure \ref{CS.vs.SS.LR} can be reached when observing the data presented in Figure \ref{CS.SLR.vs.SS.LR}. Figure \ref{CS.SLR.vs.SS.LR} shows that the $SLR_{CS}$ have a marked tendency to overestimate their $LR_{SS}$ counterparts, which may not necessarily be a problem when $H_{0_{SS}}$ is true; however, the use of $SLR_{CS}$ to report forensic evidence in court may be very detrimental to innocent suspects. We also note the particular behaviour of the relationship between $LR_{SS}$ and $SLR_{CS}$ when the variance of the control impressions is very small.

Overall, common source score-based models may be convenient to implement, but are not relevant to most examinations of forensic interest, and do not converge to the weight of forensic evidence. The lack of convergence between $LR_{SS}$ and $SLR_{CS}$ is not only a by-product of the use of a potentially non-sufficient summary statistic as the score between a pair of impressions, but also results from the inadequacy of Equation (\ref{CS.SLR}) to account for the rarity of the characteristics observed on the trace impression. 

\subsection{Suspect-centred score-based models}\label{Sus.SLR}

A second type of score-based model was proposed to be more relevant to the case at hand \citep{Helper2012, Alberink2014}. According to this type of model, the sampling distributions of $\delta(e_u,e_s)$ can be represented by the following thought experiments:
\begin{enumerate}
	\item When the prosecutor proposition is correct, $\delta(e_u,e_s)$ is a score that is calculated by comparing a trace and a control object that have been both obtained from the source of $e_s$. Thus, the sampling distribution of $\delta(e_u,e_s)$ under $H_0$ can be studied by sampling, and comparing, independent pairs of trace and control objects from the source of $e_s$.
	\item When the defence proposition is correct, $\delta(e_u,e_s)$ is a score that is calculated by comparing trace objects sampled from random sources in a relevant population to control objects from the source of $e_s$\footnote{This may seem counterintuitive, and the reader may wonder why trace objects, rather than control objects, are randomly sampled from sources in the relevant population. This sampling model is rooted in the definition of the generative model in Equation (\ref{SS.generative.model}): in the specific source scenario, there is no dispute that $e_s$ originates from the suspected source.}. 
\end{enumerate}

This type of score-based model has been designed to address the specific source pair of propositions since it is ``anchored'' on the control material obtained from the putative source. Nonetheless, it does not address the issue of the rarity of the trace characteristics as it only estimates the probability of the \textit{control material} using a sample of trace objects from the population.  Furthermore, it may be not be trivial to repeatedly sample trace and control objects from sources under controlled conditions (i.e., repeatedly resampling fingerprints from an uncooperative suspect may be tricky). To overcome the latter issue, authors have proposed to generate pseudo-fingerprints \citep{NeumannRSS, Rodriguez2012} or the use of parametric models for the score distributions \citep{Egli:2006, Egli2014}.  
%
To avoid repeatedly sampling control impressions from the donor of $e_s$, it is possible to condition the score-based model on $e_s$. The difference between the unconditioned suspect-centred score-based model described in the previous paragraph and the conditioned model is that $\delta(e_u,e_s)$ and both sampling distributions use the same fixed $e_s$. Mathematically:
\begin{eqnarray}
	SLR_{SS|e_s} & = & \frac{f(\delta(e_u,e_s),e_s|H_{0_{SS}})}{f(\delta(e_u,e_s),e_s|H_{1_{SS}})}	= \frac{f(\delta(e_u,e_s)|e_s, H_{0_{SS}})}{f(\delta(e_u,e_s)|e_s, H_{1_{SS}})}	\frac{f(e_s| H_{0_{SS}})}{f(e_s| H_{1_{SS}})}\nonumber \\	
	& = & \frac{f(\delta(e_u,e_s)|e_s, H_{0_{SS}})}{f(\delta(e_u,e_s)|e_s, H_{1_{SS}})}.
	\label{SLR.SS.es}
\end{eqnarray}

\begin{figure}
\begin{center}
\subfigure[]{\includegraphics[bb=0in 0in 11in 5.2in, scale=0.35]{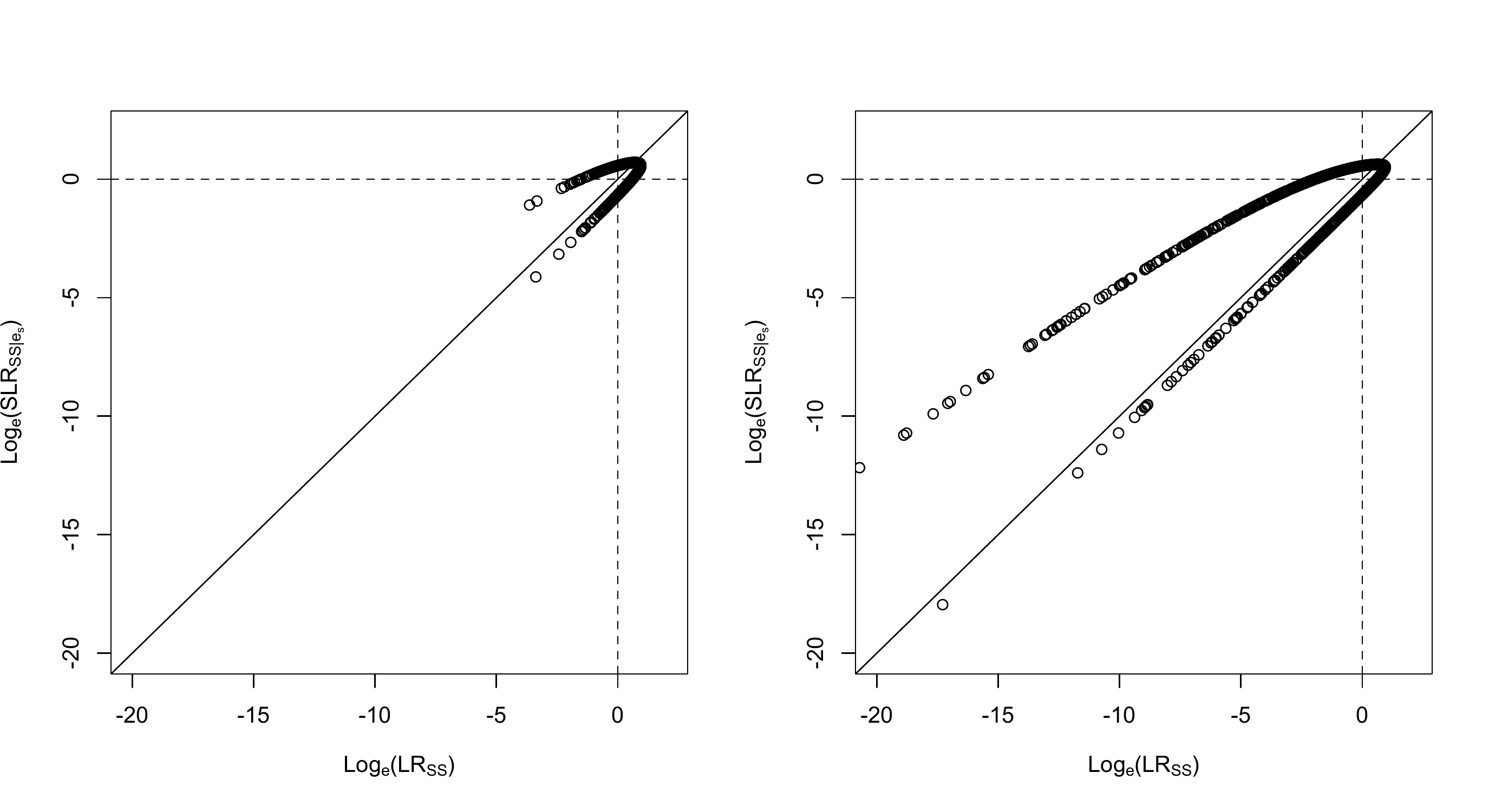}}
\subfigure[]{\includegraphics[bb=0in 0in 11in 5.2in, scale=0.35]{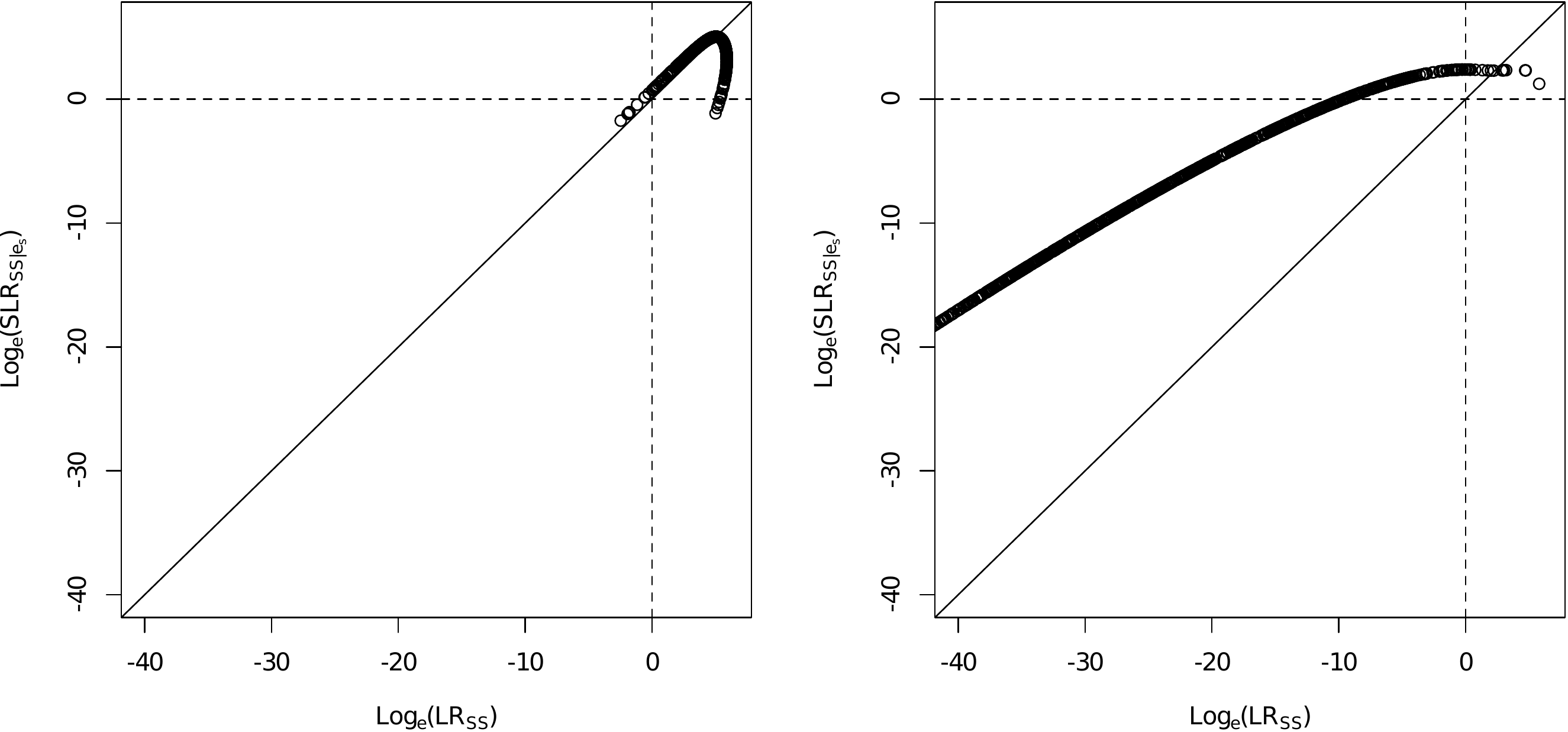}}
\subfigure[]{\includegraphics[bb=0in 0in 11in 5.2in, scale=0.35]{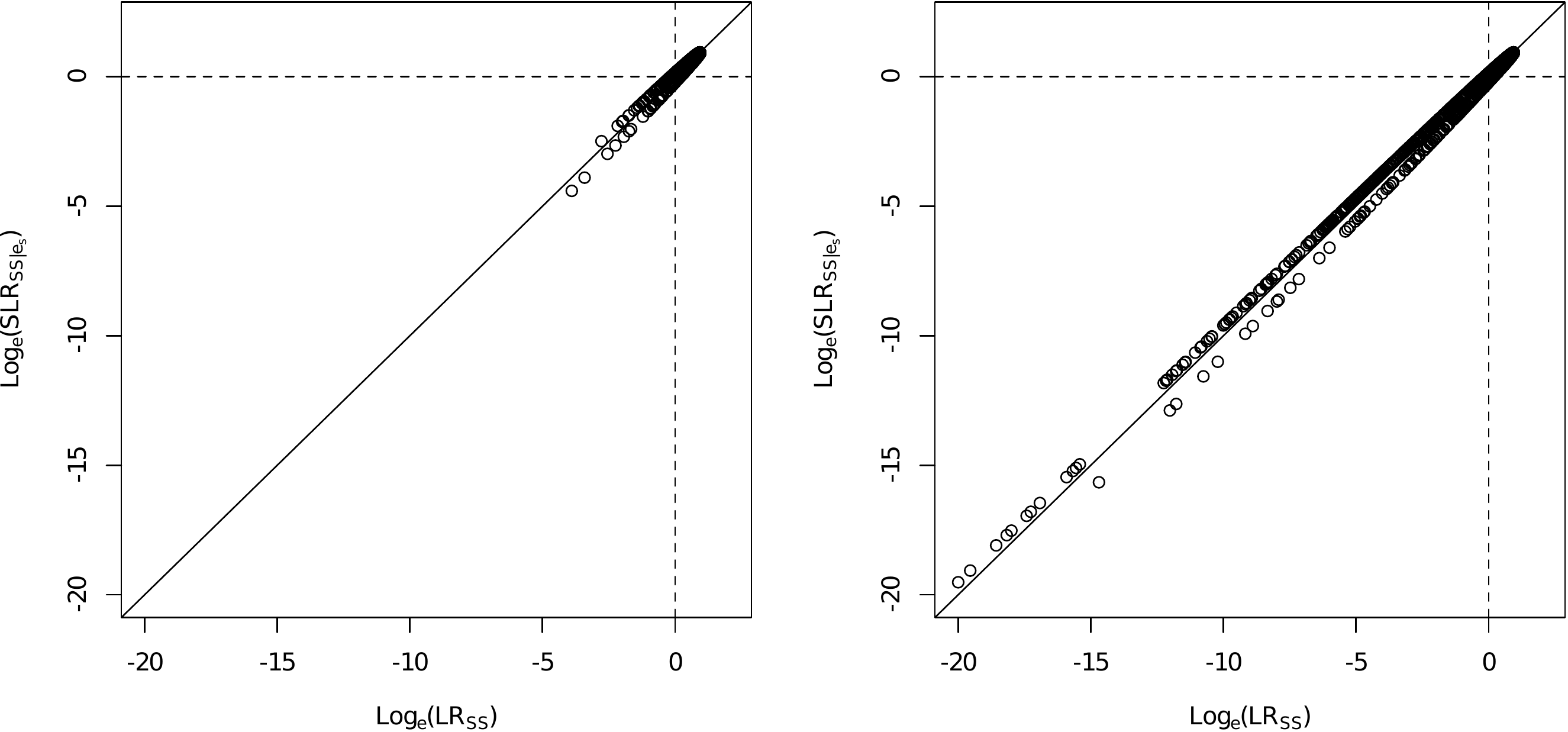}}
\end{center}
\caption{Comparisons between SLRs conditioned on $e_s$ in the specific source scenario with the LR in the specific source scenario. Columns: the left column reports the results when $e_u$ and $e_s$ have been sampled under $H_{0_{SS}}$; the right column reports the results under $H_{1_{SS}}$. Rows: (a) the source of the control object is common and has some variance; (b) the source of the control object is rare and has some variance; (c) the source of the control object is common and has virtually no variance.}
\label{SS.SLR.es.vs.SS.LR}
\end{figure}

The second ratio in Equation (\ref{SLR.SS.es}) cancels out since the characteristics of the control material have the same density irrespective of whether the source of $e_s$ is also the source of the trace material. From the generative models proposed in Equation (\ref{SS.generative.model}), and with $\delta(e_u,e_s) = (e_{u_1} - e_{u_2})^2$, we obtain the following sampling distributions for $\delta(e_u,e_s)$:
\begin{eqnarray}
	(e_u - e_s)^2 | e_s, H_{0_{SS}}  & \sim & \frac{1}{\sigma^2_u}\chi^2 \left( \frac{(e_u - e_s)^2}{\sigma^2_u}, \lambda=\frac{(\mu_p-e_s)^2}{\sigma^2_u} \right)\nonumber \\
	(e_u - e_s)^2 | e_s, H_{1_{SS}}  & \sim & \frac{1}{\sigma^2_u + \sigma^2_p}\chi^2 \left( \frac{(e_u - e_s)^2}{\sigma^2_u + \sigma^2_p}, \lambda = \frac{(\mu - e_s)^2}{\sigma^2_u + \sigma^2_p} \right).
\end{eqnarray}
These sampling distributions enable us to compare the likelihood ratio of interest, $LR_{SS}$, with its proxy, $SLR_{SS|e_s}$. This comparison is reported in Figure \ref{SS.SLR.es.vs.SS.LR} using the same parameter values as those used to generate the results presented in Figures \ref{CS.vs.SS.LR} and \ref{CS.SLR.vs.SS.LR}\footnote{Note that the results presented in Figure \ref{SS.SLR.es.vs.SS.LR} are highly dependent of the value chosen for $e_s$. In particular, the patterns in Figures \ref{SS.SLR.es.vs.SS.LR}(a) and (b) are very sensitive to the value of $e_s$.}.

The model proposed in Equation (\ref{SLR.SS.es}) certainly seems to be a reasonable ad-hoc solution: it is specific to the suspected source; the required sampling/simulation of trace objects from relevant sources can be achieved by using a suitable parametric model of the score distributions; furthermore, under the reasonable assumption, for some evidence types, that control objects have very limited variance, it will mostly converge to the specific source likelihood ratio of interest (Figure \ref{SS.SLR.es.vs.SS.LR}(c)). Unfortunately, this type of model is plagued by a fundamental lack of coherence: indeed, with these models, a given piece of evidence can provide support for either of the alternative propositions, depending on which proposition is considered first. 

To demonstrate this lack of coherence, consider a model designed to address the two following specific source propositions:
\begin{itemize}
	\item[] $H_{A}$: $e_{u}$ originates from Source A; 
	\item[] $H_{B}$: $e_{u}$ originates from Source B.
\end{itemize}
We observe the trace object as well as two control objects, one from Source A and one from Source B. The specific source generative models under $H_A$ and $H_B$ are described below.\\ 
Under $H_A$, we have:
\begin{align*}
e_a & = \mu_a + a, \text{where} \ a \sim N(0,\sigma^2_a); \hspace{1cm} e_u = \mu_a + u, \text{where} \ u \sim N(0,\sigma^2_u);\\
e_b & = \mu_b + b, \text{where} \ b \sim N(0,\sigma^2_b).
\end{align*}
Under $H_B$, we have:
\begin{align*}
e_a & = \mu_a + a, \text{where} \ a \sim N(0,\sigma^2_a);\\
e_b & = \mu_b + b, \text{where} \ b \sim N(0,\sigma^2_b); \hspace{1cm} e_u = \mu_b + u, \text{where} \ u \sim N(0,\sigma^2_u).
\end{align*}
The specific source likelihood ratio that addresses $H_A$ and $H_B$ is:
\begin{equation}
	LR_{SS} = \frac{f(e_u,e_a,e_b|H_{A})}{f(e_u,e_a,e_b|H_{B})}	 = \frac{f(e_u|H_{A})}{f(e_u|H_{B})} = \left(\frac{f(e_u|H_{B})}{f(e_u|H_{A})} \right)^{-1}.
\end{equation}
Thus, $LR_{SS}$ coherently supports the same proposition irrespective of which one is considered first. However, the specific source $SLR_{SS}$ conditioned on the control impression considered by the first proposition is:
\begin{equation}
	SLR_{SS|e_a} = \frac{f(\delta(e_u,e_a)|e_a, H_{A})}{f(\delta(e_u,e_a)|e_a, H_{B})}	 \neq \left(SLR_{SS|e_b}\right)^{-1} =\left(\frac{f(\delta(e_u,e_b)|e_b, H_{B})}{f(\delta(e_u,e_b)|e_b, H_{A})} \right)^{-1}.
	\label{SLR.SS.not.coherent}
\end{equation}
Equation (\ref{SLR.SS.not.coherent}) shows that $SLR_{SS|e_s}$ is not coherent in general since it potentially does not support the same proposition depending on which one is considered first. This lack of coherence can similarly be demonstrated for the unconditioned  suspect-centred score-based likelihood ratio.

\begin{figure}[h]
\begin{center}
\includegraphics[width=\textwidth]{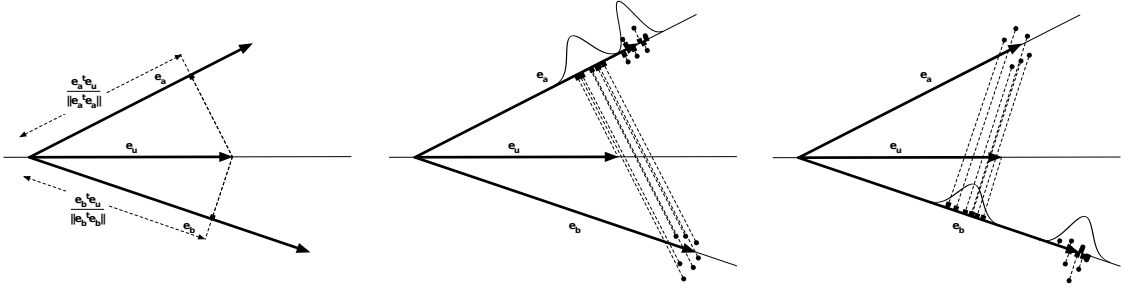}
\caption{Left panel: vector projection of $\textbf{e}_u$ onto $\textbf{e}_a$ and $\textbf{e}_b$. Middle panel: vector projection of all (pseudo-)trace objects, including $\textbf{e}_u$, onto $\textbf{e}_a$. Right panel: vector projection of all (pseudo-)trace objects, including $\textbf{e}_u$, onto $\textbf{e}_b$.}
\label{vectors.SLR.2}
\end{center}
\end{figure}

The conditioning of $\delta(e_u,e_s)$ on $e_s$ has an interesting geometric interpretation. When $e_s$ is fixed, all trace objects from the source considered under $H_{0_{SS}}$ and from the sources from the population considered under $H_{1_{SS}}$ are compared to the same control object. It is thus possible to consider that all scores considered in our thought experiment are equivalent to the scalar projections of the vectors representing all trace objects onto a vector space defined by $e_s$. Figure \ref{vectors.SLR.2} illustrates this interpretation. 

In Figure \ref{vectors.SLR.2}, a vector $\textbf{e}_u$, representing a trace object recovered in connection with a crime, is compared to objects from sources $A$ and $B$, represented by two mean vectors, $\textbf{e}_a$ and $\textbf{e}_b$. The left panel shows the orthogonal projection of $\textbf{e}_u$ onto $\textbf{e}_a$ and $\textbf{e}_b$. The resulting scalars are the scores calculated by $\delta(\textbf{e}_u,\textbf{e}_a)$ and $\delta(\textbf{e}_u,\textbf{e}_b)$, which are, in this case, equivalent to $\kappa(\textbf{e}_u,\textbf{e}_a)$ and $\kappa(\textbf{e}_u,\textbf{e}_b)$. 

The middle panel shows the orthogonal projections of (1) $\textbf{e}_u$, (2) multiple pseudo-trace objects sampled from the source of $\textbf{e}_b$ (represented by dots near the tip of $\textbf{e}_b$), and (3) multiple pseudo-trace objects sampled from the source of $\textbf{e}_a$ (represented by dots near the tip of $\textbf{e}_a$) onto  $\textbf{e}_a$. The two density functions represent the distributions of the scalar projections of these vectors onto $\textbf{e}_a$. According to the middle panel, $SLR_{SS|e_{a}}$, in Equation (\ref{SLR.SS.not.coherent}), is equivalent to the ratio of the likelihoods of the scalar projection of $\textbf{e}_u$ onto $\textbf{e}_a$ evaluated using the two distributions of the scalar projections of the trace objects from both suspects onto $\textbf{e}_a$. We see that, in this case, $SLR_{SS|e_{a}}$ would support the proposition that $\textbf{e}_u$ was made by Source $B$. The right panel shows the same information as in the middle panel, but this time, projected onto $\textbf{e}_b$. We see that in this case, $SLR_{SS|e_{b}}$ would support the proposition that  $\textbf{e}_u$ was made by Source $A$. This geometric interpretation holds in the general case, when the alternative hypothesis is not specific to a single donor, but considers a population of sources as in Section \ref{SS.Hypotheses}. In the general case, all pseudo-trace objects from all sources of the relevant population are projected onto the vector representing a single source under $H_{0_{SS}}$ or under $H_{1_{SS}}$. This results in the same lack of coherence in the support of the evidence for alternative propositions representing different sources.\\

\subsection{Trace-centred score-based models}

The two types of models presented above lack the ability to account for the rarity of the characteristics of the trace object, which is crucial to properly quantify the weight of forensic evidence. To remedy this shortcoming, the use of trace-centred score-based models is found in \cite{Alberink2014}. This type of model is somewhat similar to the family of suspect-centred models. However, since it is not possible to sample more trace and control objects from the true source of $e_u$ (since it is unknown), these models must be conditioned on the observed trace, $e_u$. Mathematically, we have: 
\begin{eqnarray}
	SLR_{SS|e_u} & = & \frac{f(\delta(e_u,e_s),e_u|H_{0_{SS}})}{f(\delta(e_u,e_s),e_u|H_{1_{SS}})}	= \frac{f(\delta(e_u,e_s)|e_u, H_{0_{SS}})}{f(\delta(e_u,e_s)|e_u, H_{1_{SS}})}	\frac{f(e_u| H_{0_{SS}})}{f(e_u| H_{1_{SS}})}\nonumber \\	
	& = & \frac{f(e_u| H_{0_{SS}})}{f(e_u| H_{1_{SS}})}.
	\label{SLR.SS.eu}
\end{eqnarray}
Interestingly, the second ratio in Equation (\ref{SLR.SS.eu}) does not cancel out. Indeed, the likelihood of observing the trace object, $e_u$, is very different under $H_{0_{SS}}$ and $H_{1_{SS}}$. In fact, this ratio corresponds exactly to the likelihood ratio of interest presented in Equation \ref{SS.LR}. On the contrary, the first ratio, which includes the score, does cancel since $e_u$ is fixed under both propositions due to conditioning, and $e_s$ has the same distribution under both propositions in the specific source scenario. This can be seen when using the generative model in Equation (\ref{SS.generative.model}) with $\delta(e_u,e_s) = (e_{u_1} - e_{u_2})^2$, which results in the same sampling distributions under both propositions:
\begin{eqnarray}
	(e_u - e_s)^2 | e_u, H_{0_{SS}}  & \sim & \frac{1}{\sigma^2_s}\chi^2 \left( \frac{(e_u - e_s)^2}{\sigma^2_s}, \lambda=\frac{(e_u - \mu_p)^2}{\sigma^2_s} \right);\nonumber \\
	(e_u - e_s)^2 | e_u, H_{1_{SS}}  & \sim & \frac{1}{\sigma^2_s}\chi^2 \left( \frac{(e_u - e_s)^2}{\sigma^2_s}, \lambda=\frac{(e_u - \mu_p)^2}{\sigma^2_s} \right).
	\label{sampling.SLR.SS.eu}
\end{eqnarray}
The results in Equations (\ref{SLR.SS.eu}) and (\ref{sampling.SLR.SS.eu}) may seem suspicious at first. Some readers may consider that, under $H_{1_{SS}}$, the sampling distribution should involve control objects from sources in the relevant population. However, $H_{1_{SS}}$ is very clear on the origins of $e_s$: its source is undisputed and it is the same specific source considered in $H_{0_{SS}}$ (see Section \ref{SS.Hypotheses} and Equation (\ref{SS.generative.model})). Geometrically, this type of model has a similar interpretation as the suspect-centred model. The trace-centred model can be understood as the projection of all control objects onto a vector space defined by $e_u$. However, as mentioned above, the only control objects available in this type of model are control objects of the source of $e_u$ under both alternative propositions. Therefore, the first ratio in $SLR_{SS|e_u}$ will always be one. In conclusion, it appears that trace-centred score-based likelihood ratios are not very useful.  

\subsection{Asymmetric score-based models}
\label{AsySLR}

A last type of models, which historically happened at the very early stages of the development of score-based likelihood ratios and seems to be the most prevalent in the literature \citep{Champod2001.Ear, Meuwly2001, Meuwly2006,  Egli:2006, Gonzalez-Rodriguez:2003, Gonzalez-Rodriguez:2005, Gonzalez:2006, Neumann:2007, Neumann2009.inks}, focuses on the putative source in its numerator and on some measure of the rarity of the characteristics of the trace object in the denominator.

According to this type of model, the sampling distributions of $\delta(e_u,e_s)$ can be represented by the following thought experiments:
\begin{enumerate}
	\item When the prosecutor proposition is correct, $\delta(e_u,e_s)$ is a score that is calculated by comparing a trace and a control object that have been both obtained from the source of $e_s$. Thus, the sampling distribution of $\delta(e_u,e_s)$ under $H_0$ can be studied by sampling, and comparing, independent pairs of trace and control objects from the source of $e_s$. This experiment is similar to the one described for the numerator of the suspect-centred score-based likelihood ratio in Section \ref{Sus.SLR}. It may or may not be conditioned on $e_s$. 
	\item When the defence proposition is correct, $\delta(e_u,e_s)$ is a score that is calculated by comparing the observed trace objects to a random control object in the population of potential sources. Thus, the sampling distribution of $\delta(e_u,e_s)$ under $H_1$ can be studied by sampling control objects from the sources in the population of potential sources and comparing them to $e_u$. This experiment is somewhat similar to the one that is done to address the denominator of the common source score-based likelihood ratio, with the modification that control objects are sampled from the population of sources (instead of trace objects in Section \ref{CS.SLR.section}) and that the denominator is conditioned on the observed trace, $e_u$.
\end{enumerate}

Mathematically, the asymmetric score-based likelihood ratio could be represented as: 
\begin{eqnarray}
	SLR_{ASY} & = & \frac{f(\delta(e_u,e_s),e_s|H_{0_{SS}})}{f(\delta(e_u,e_s),e_u|H_{1_{CS}})}	= \frac{f(\delta(e_u,e_s)|e_s, H_{0_{SS}})}{f(\delta(e_u,e_s)|e_u, H_{1_{CS}})}	\frac{f(e_s|H_{0_{SS}})}{f(e_u|H_{1_{CS}})}.
	\label{SLR.ASY}
\end{eqnarray}

While the first ratio in the right-hand part of Equation \ref{SLR.ASY} seems appealing, at first, in the sense that it is both suspect and trace anchored, it is clear from Equation \ref{SLR.ASY} that the ratio does not consider the same evidence in the numerator and in the denominator, which is a logical violation of the concept of likelihood ratio. Furthermore, the second ratio in the right-hand part of Equation \ref{SLR.ASY} does not cancel. Thus, we do not see that $SLR_{ASY}$ can possibly converge to the specific source Bayes factor of interest. 

\section{Discussion and conclusion}

Various attempts have been made to quantify the weight of fingerprint evidence. Most of these attempts suffer from severe shortcomings, which result in unpredictable bias with respect to the Bayesian inference framework. Some of these shortcomings include addressing the common source scenario instead of being specific to a suspected donor, failing to account for the rarity of the features observed on the latent impression, or providing incoherent evidence which may support both of two mutually exclusive propositions. 

A Bayes factor is the ratio between two probabilities. Following \cite{Good:1950}, \cite{Jeffreys:1961}, \cite{Savage:1972}, \cite{Jaynes:2003}, \cite{Lindley:2006} and many others (for a recent review see \citep{Taroni:2016}), we took the view throughout this paper that probabilities can only represent the degree of belief of an individual about an event and are influenced by the information that he has about the event. Two individuals considering a particular event from two different perspectives may very well have different degrees of belief about that event. Thus, probabilities are subjective in the sense that they represent the personal relationship between the subject and the event. 

The Bayes factor is not an intrinsic property of the evidence in itself, and we want to be very clear that we do not claim that there is such thing as a true or universal Bayes factor for a given piece of evidence. Different scientists may assign different weights to forensic evidence if they characterise the evidence material using different types of features or measure it using different analytical techniques\footnote{For example, glass fragments may be characterised by their refractive index or by their elemental composition.}, if they organise the data in different ways\footnote{\cite{Neumann2015} describe a method to characterise the spatial relationships between fingerprint landmarks (i.e., minutiae) using triangles and used these triangles to assign probability distributions to minutiae constellations. However, it is certainly possible to characterise the spatial relationships between minutiae in many other ways.} or if they choose different models to represent the data\footnote{Given a set of observations, a scientist may choose to rely the assumption that the data are normally distributed, use another parametric model, or use non-parametric models.}.

Nonetheless, the adjectives subjective or personal are not meant to suggest, or justify, that probabilities can be assigned arbitrarily, or reflect sloppy thinking \citep{Lindley:2006, Taroni:2016}. Bayes factors have fundamental properties, which should be satisfied by any method designed to quantify the weight of forensic evidence. These general properties are applicable to any model designed to assign Bayes factors.

By definition, a Bayes factor provides two pieces of information: which one of the two competitive propositions is favoured by the evidence, and the amount of support provided by the evidence. While it may be relatively easy to test the accuracy of a probabilistic model in large scale simulation settings (i.e., whether the correct proposition is consistently favoured \citep{Neumann:2007, Haraksim2015, Leegwater2017}), determining the appropriateness of the amount of support is an open problem. 

Concentrating on the accuracy of a probabilistic model is arguably equivalent to considering the method as a deterministic decision-making engine with known error rates. The use of this type of techniques to infer the source of forensic traces has been explicitly discouraged by some \citep{ENFSI2016} but encouraged by others \citep{NAS:2009, PCAST2016}. We agree with \cite{Champod2015}, \cite{Evett2017} and \cite{Morrison} in that error rates are only an average measure of performance over a population and do not provide information regarding the support of the evidence in individual cases. 

Accuracy does not inform on whether a particular method supports a given proposition with the appropriate magnitude. Yet, in the legal context, the magnitude of the Bayes factor is critical. Grossly under- or overestimating the weight of the evidence can seriously distort the fact-finding process and be prejudicial to the accused\footnote{Consider that a partial fingerprint is recovered at a crime scene and is compared to the friction ridge skin of an accused. A jury will perceive the probative value of the evidence differently and may reach different conclusions if the reported Bayes factor is one thousand, or one billion. Depending on the case circumstances, the defence may be able to argue that the other evidence against the defendant is sufficiently weak that a Bayes factor of one thousand is not sufficient to reach a conclusion beyond reasonable doubt. A similar argument will be excessively difficult to make if the reported Bayes factor for the forensic evidence turns out to be one billion. }. Some authors have proposed methods to study the magnitude of the values outputted by probabilistic models \citep{Brummer2006, Ramos2013b, Ramos2013, Haraksim2015, Leegwater2017}. We believe that some of these methods have merit and we will discuss them in a future paper. However, none of these methods answers the question of the appropriateness of these magnitudes or addresses the soundness of the scientific foundations of a given probabilistic model.

It is also always possible to consider that the magnitude of the value produced by the model is important as a rank statistic, but not as a value in itself. In this case, we fall within the realm of likelihoodist inference \citep{Royall:1997, Kaye:2012}, which is not necessarily compatible with Bayesian inference. 

Therefore, our conclusion is that none of the score-based models proposed to date can be considered as suitable proxies of the Bayes factor of interest. We are not arguing that these ad-hoc methods are not useful in their own way, but the harsh reality is that if one wants to abide by the idea that forensic evidence should be reported within a Bayesian paradigm, then one cannot use score-based likelihood ratios. We appreciate that the use of scores may be the only viable method to reduce the complexity of forensic evidence, but more efforts should be spent in the development of more rigorous models for handling these scores (see \citep{Armstrong:2017, Ausdemore.2stages.2019, Hendricks2019ABC} for some early work on these models). 

\bibliographystyle{chicago}

\end{document}